\documentclass[conference]{IEEEtran}
\IEEEoverridecommandlockouts

\usepackage{cite}
\usepackage{amsmath,amssymb,amsfonts}
\usepackage{algorithmic}
\usepackage{graphicx}
\usepackage{textcomp}
\usepackage{xcolor}

\usepackage{booktabs}
\usepackage{multirow}
\usepackage{makecell}
\usepackage{stfloats}
\usepackage{hyperref}

\def\BibTeX{{\rm B\kern-.05em{\sc i\kern-.025em b}\kern-.08em
    T\kern-.1667em\lower.7ex\hbox{E}\kern-.125emX}}
\begin{document}

\title{Architecting Long-Context LLM Acceleration with Packing-Prefetch Scheduler and Ultra-Large Capacity On-Chip Memories\\

\thanks{Submitted to IEEE MICRO Special Issue "AI for Hardware and Hardware for AI" for review.}
}

\author{
    Ming-Yen Lee$^{\dagger}$, Faaiq Waqar$^{\dagger}$, Hanchen Yang$^{\dagger}$, Muhammed Ahosan Ul Karim$^{*}$, Harsono Simka$^{*}$, Shimeng Yu$^{\dagger}$ \\
    $^{\dagger}$\textit{Georgia Institute of Technology, Atlanta, GA, 30332, USA} \\
    $^{*}$\textit{Samsung Semiconductor Inc., San Jose, CA, 95134, USA}
}

\maketitle

\begin{abstract}
Long-context Large Language Model (LLM) inference faces increasing compute bottlenecks as attention calculations scale with context length, primarily due to the growing KV-cache transfer overhead that saturates High Bandwidth Memory (HBM). While prefetching techniques mitigate cache misses by fetching KV data in advance, their spatial and temporal benefits present new opportunities to exploit. This work proposes a packing-prefetch scheduling architecture with monolithic 3D (M3D) back-end-of-line (BEOL) compatible embedded memories with ultra-large on-chip capacity to accelerate long-context LLM inference. Our optimizations demonstrate 8.06× decode speedup and 1.83× overall latency reduction on Llama3.1-8B using TPUv6e-like hardware with additional 512MB BEOL memories over the serial execution. Evaluations of multi-request workloads on TPU-like architectures show 1.7×–2.4× throughput improvement and 1.5×–2.4× HBM bandwidth reduction compared to packing-only methods on Llama3.1-8B and Llama3.1-70B models. With the co-design of packing, prefetching, and BEOL memories, our approach alleviates HBM constraints and enables efficient long-context LLM inference.
\end{abstract}


\section{Introduction}
\begin{figure}
\centerline{\includegraphics[width=21pc]{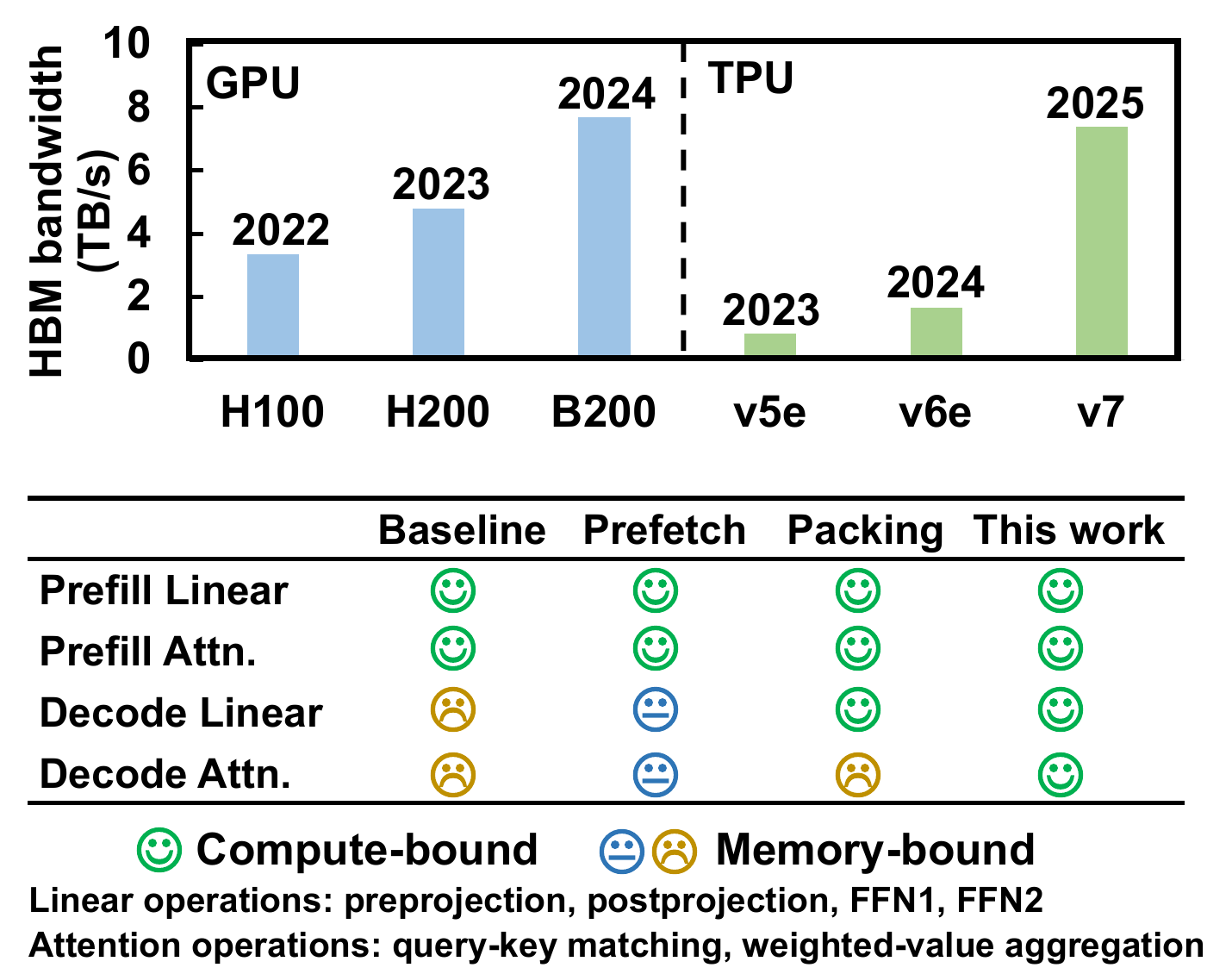}}
\caption{HBM bandwidth trends in GPUs and TPUs, and the breakdown of compute-bound and memory-bound operations during the two phases of LLM inference. Memory-bound operations impose high bandwidth pressure on HBM.}\vspace*{-5pt}
\label{intro}
\end{figure}

{L}arge Language Models (LLMs) have demonstrated remarkable performance across diverse applications, including dialogue systems, code synthesis, and summarization. The rapid adoption of LLM has made efficient inference a critical challenge, especially on reducing latency metrics such as time-to-first-token and time-between-tokens, to meet the demands of real-time applications. Prior works have explored various algorithmic optimizations and hardware accelerations. With batching and operator fusion\cite{fusion}, the prefill phase of LLM inference has become compute-bound with reduced off-chip data transfers and high data reuse. This means prefill performance is mainly limited by compute throughput rather than HBM bandwidth. However, a key bottleneck remains in the decode phase, where auto-regressive token generation suffers from high data transfer overhead between off-chip HBM and on-chip compute resources. As shown in Figure \ref{intro}, even with the improved HBM bandwidth, decode latency remains constrained by KV-cache transfer, especially as compute units continue to advance. This challenge is exacerbated by the growing context lengths in modern LLMs, which increase KV-cache sizes and place even greater pressure on HBM bandwidth during attention computations.

The decode phase primarily consists of two operations: linear operations and attention operations. Linear layers can be optimized by packing compute-bound operations from other requests' prefill phase since model weights are shared across requests\cite{sarathi1}. Unfortunately, attention operations still require separate processing and remain memory-bound. Prefetching techniques have been proposed as a complementary solution. For example, prior work \cite{prefetch} leveraged KV-cache and weight prefetching during GPU communication overlaps. We believe that the full temporal and spatial potentials of prefetching can be further explored. Firstly, additional prefetching time can be exploited by overlapping with the compute-bound prefill phase, instead of only prefetching during the communication interval. Second, M3D memories\cite{cmos+x} have emerged as a promising solution for high-density, high-speed on-chip buffer, enabling more prefetched KV-cache accommodation.

In this work, we propose a packing-prefetch scheduling architecture with ultra-large capacity on-chip memories to alleviate HBM bottlenecks and improve long-context LLM inference efficiency. Temporally, we utilize the residual HBM bandwidth during compute-bound operations by interleaving prefill and decode phases, providing sufficient time for KV-cache data prefetching. Spatially, we introduce high-density M3D embedded memories to satisfy the capacity requirement for prefetched data storage. To systematically evaluate our approach, we develop an LLM simulation framework by customizing Timeloop\cite{timeloop}, modeling end-to-end inference on AI accelerators. Evaluation results show that our packing-prefetch scheduling achieves up to 8.06× token-to-token decode speedup and 1.83× overall latency reduction on Llama3.1-8B performed by TPUv6e-like accelerators with an additional 512MB M3D memories compared to serial execution. We also conduct service-level evaluations with multi-request workloads on the openchat\_sharegpt4 and arxiv\_summarization datasets. On TPU-like architectures, our method achieves 1.7×–2.4× throughput improvements and 1.5×–2.4× HBM bandwidth reduction compared to packing-only baselines across Llama3.1-8B and Llama3.1-70B models.

\section{Background}
\begin{figure}
\centerline{\includegraphics[width=21pc]{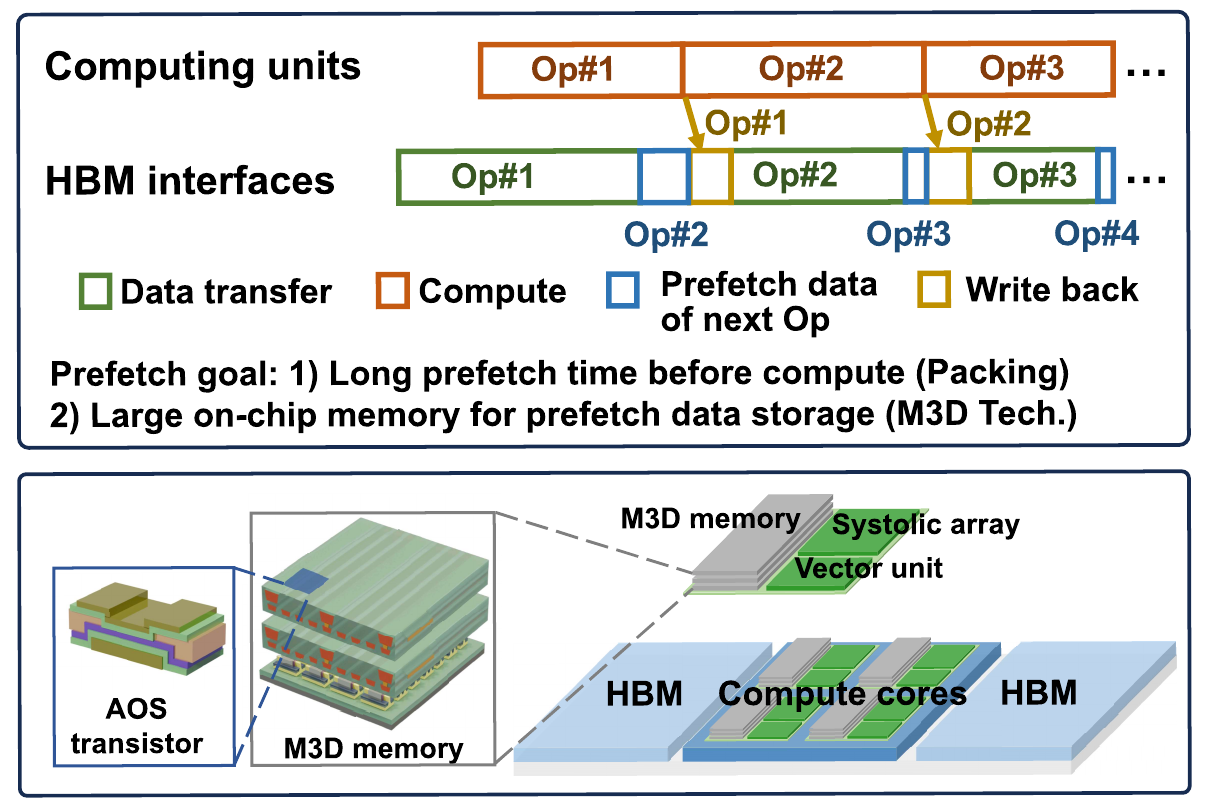}}
\caption{Time diagram of computing units and HBM data transfer on AI accelerators. Packing and high-density M3D technologies provide the chance to maximize the benefit of prefetching.}\vspace*{-5pt}
\label{background}
\end{figure}

Figure \ref{background} illustrates the time diagram of compute units and HBM data transfer during multi-operation execution on TPU-like architectures. Typically, TPU accelerators consist of high-performance compute units (e.g., systolic arrays and vector units), on-chip memory for intermediate operands, and off-chip HBM that stores the full model and KV-cache. Computation is initiated after a partition of operands is fetched from HBM into the on-chip memory. During execution, the compute units and HBM operate in a pipelined manner. While computations are performed, data can be transferred on- and off-chip simultaneously through double-buffering mechanisms in on-chip memory. For compute-bound operations, computation time dominates over data transfer time, leaving residual HBM bandwidth that can be leveraged to prefetch model weights or KV-cache data for the upcoming operation. After completion, outputs are either written back to HBM or remain on-chip for the next operation, which is known as operator fusion\cite{fusion}. Consequently, subsequent operations benefit from reduced data transfer latency due to the prefetching in previous compute-bound phases.

To fully exploit prefetching, which mitigates compute stalls for the next operation and alleviates HBM bandwidth bottlenecks, particularly for memory-bound attention operations in the decode phase, our approach focuses on two targets: sufficient prefetching time and prefetching space. First, borrowing HBM bandwidth from linear operations in the decode phase is often ineffective, since these operations can easily become memory-bound. Even though large batch size improves weight reuse, it also increases inference latency and KV-cache footprint for HBM. In contrast, prefill-phase operations tend to be compute-bound and can easily saturate compute resources\cite{sarathi1}. By interleaving prefill and decode phases across different request batches, we can utilize the residual HBM bandwidth during prefill to prefetch KV-cache data required for subsequent decode attention.

Second, to take advantage of this extended prefetching window, sufficient on-chip memory capacity is critical to store prefetched data. M3D embedded memories have emerged as a promising solution to address this storage demand. Leveraging recent advances in the device technologies that are being pursued by leading foundries\cite{samsungigzo,tsmc1t1c}, amorphous oxide semiconductor (AOS) transistors enable low-leakage and high-speed embedded DRAM-like on-chip buffers. As shown in Figure \ref{background}, dense memory cells (e.g., two-transistor gain cells) are stacked above the logic layer, with peripheral circuits fabricated in the front-end-of-line (FEOL) and multi-layer memory arrays integrated using the BEOL interconnect. This CMOS+X approach could potentially achieve several hundreds of megabytes of on-chip memory capacity \cite{cmos+x}, providing the space for storing prefetched KV-cache data. As a reference, state-of-the-art V-cache by bonding SRAM on top of microprocessor's core could achieve only 96MB capacity\cite{amdvcache}. By combining prefetching scheduling with packing and high-density M3D memories, our proposed packing-prefetch scheduling architecture optimizes the prefetching opportunity for attention computation in the decode phase. Both prefill and decode phases become compute-bound, which alleviates HBM bandwidth bottlenecks and improves the efficiency of long-context LLM inference.

\section{Proposed Packing-Prefetch Scheduling}
\begin{figure}
\centerline{\includegraphics[width=21pc]{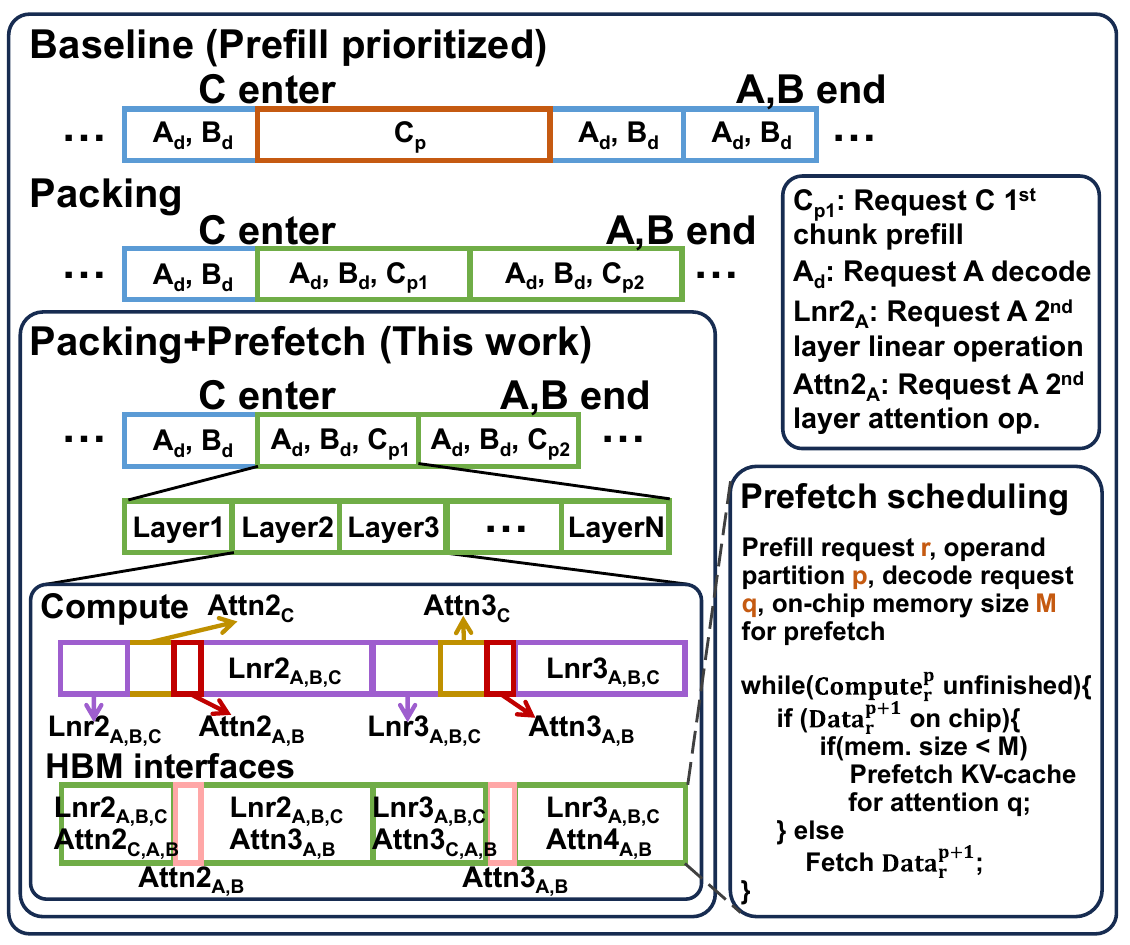}}
\caption{Time diagram and algorithm of proposed scheduling.}\vspace*{-10pt}
\label{scheduler}
\end{figure}

Figure \ref{scheduler} illustrates the time diagram and operational principle of proposed packing-prefetch scheduling, which combines inter-request packing and KV-cache prefetching. In the baseline approach, requests A and B are in the decode phase when a new request C arrives. The scheduler initiates request C’s prefill phase and queues all three requests for the next decode cycle. With packing, request C is split into two chunks, each interleaved with the decode operations of requests A and B\cite{sarathi2}. This interleaving of prefill and decode phases overlaps compute-bound linear operations from the prefill phase of request C with memory-bound linear operations from the decode phase of A and B. Therefore, the weight reuse is improved, converting the decode linear operations from memory-bound to compute-bound. However, even with packing, attention operations in the decode phase remain memory-bound, dominated by the transfer of large KV-cache data from HBM. To address this bottleneck, we introduce prefetching during compute-bound phases, leveraging residual HBM bandwidth by 
interleaving the data fetching of current packed linear operation and KV-cache data prefetching for the upcoming attention calculation.

As shown in the time diagram, operations are scheduled in a layer-by-layer sequence. While compute units execute packed linear operations, the scheduler dynamically allocates HBM bandwidth for data fetching guided by two conditions: (1) Operand fetch priority: If operand partitions for the current linear operation are not loaded into on-chip memory, HBM prioritizes their transfer. (2) KV-cache prefetch opportunity: If those operands are already on-chip and there is available on-chip memory, the scheduler uses the idle HBM bandwidth to prefetch KV-cache data for the next attention operation. For example, consider two consecutive layers (e.g., Layer 2 and Layer 3). During the linear operation before attention in Layer 2, the compute units process the packed prefill/decode operations of requests A, B, and C, while HBM simultaneously prefetches the KV-cache for the upcoming attention in Layer 2 for requests A and B. During attention execution of A and B, the required KV-cache is already on-chip, eliminating HBM transfer delays. Similarly, during the linear operations after attention in Layer 2 and before attention in Layer 3, the scheduler prefetches KV-cache for Layer 3, depending on HBM bandwidth availability and memory space as discussed in the above two conditions. By overlapping compute-bound operations with KV-cache prefetching, our packing-prefetch scheduling eliminates HBM-induced latency stalls during attention, transforming operation stages from memory-bound to compute-bound. As a result, all inference stages become compute-bound with high hardware utilization, improving inference throughput and relieving the pressure for HBM bandwidth.

\section{Proposed LLM Inference Framework and Architecture}
\begin{figure}
\centerline{\includegraphics[width=21pc]{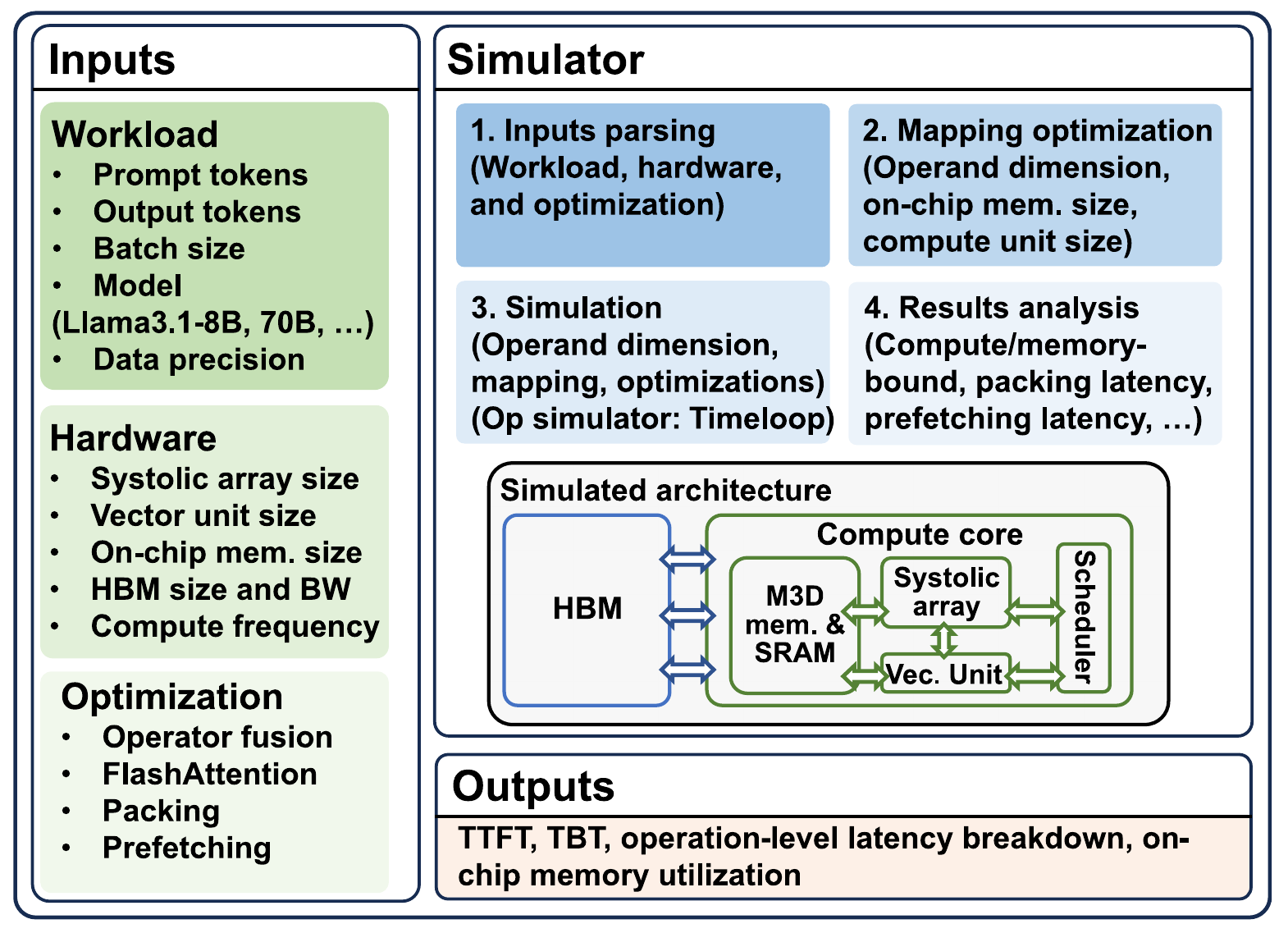}}
\caption{Overview of proposed simulation framework.}\vspace*{-10pt}
\label{framework}
\end{figure}
To systematically model and evaluate the performance of LLM inference under different optimization strategies, we develop an end-to-end simulation framework that supports configurable LLM models, hardware architectures, and optimization techniques. Figure \ref{framework} shows the overview of proposed framework, which accepts three primary inputs: 1) LLM workload, including parameters such as input/output token lengths, batch size, LLM models (e.g., Llama3.1-8B, Llama3.1-70B), and data precision; 2) Hardware architecture, specifying the number of compute resources (e.g., systolic arrays, vector units), their operating frequency, and memory hierarchy including on-chip memories and off-chip HBM; 3) Optimization techniques, such as operator fusion, FlashAttention\cite{flashattention}, packing\cite{sarathi2}, and prefetching\cite{prefetch}. 

After receiving the configurations, the simulator parses all parameters and performs operation-level mapping based on operand dimensions, available memory, and compute resources. A mapping space exploration using iterative search then identifies an optimized mapping strategy. Note that the operand parameters in linear and attention operations are smaller than in typical convolutional layers, resulting in a narrower search space and a runtime of under one minute for overall mapping optimization. Each optimization technique is implemented either during the mapping phase or into the configuration files for the external simulator. Operator fusion is applied automatically when operand dimensions, on-chip memory capacity, and mapping constraints permit. FlashAttention is implemented through head-level tiling, and is also compatible with the operator fusion between query-key matching and value-weighted aggregation for off-chip data transfer reduction. Packing and prefetching are directly modeled by specifying operation dimensions and evaluating their effects on execution overlaps and data movement.

With all configurations set, we invoke the Timeloop simulator\cite{timeloop} to perform operation-level simulation. For each operation, the framework records compute latency, data transfer time, and on-chip memory utilization. The final report provides key inference metrics, including Time-to-First-Token (TTFT) for prefill phase and Time-between-Tokens latency (TBT) for decode phase, both with and without optimization techniques (e.g., packing and prefetching). We can compared these metrics quantitatively between our scheduling scheme and other optimization baselines.

\begin{table}[t]
\caption{Configurations of LLMs and TPU-like architectures.}
\label{tab:llm_hw_config}
\centering
\renewcommand{\arraystretch}{1.7}  
\resizebox{\linewidth}{!}{
\begin{tabular}{lcc}
\toprule
\textbf{LLM model} & \textbf{Llama3.1-8B\cite{llama}} & \textbf{Llama3.1-70B\cite{llama}} \\
\midrule
\makecell[l]{TPU-like\\architecture} & TPUv6e-like & TPUv7-like \\
\makecell[l]{FP16\\performance} & 918 TFLOPS & 4614 TFLOPS \\
Systolic array & 128$\times$128$\times$16 & 256$\times$256$\times$16 \\
Vector unit & 128$\times$16$\times$16 & 256$\times$32$\times$16 \\
\makecell[l]{On-chip\\memory} & \makecell{80MB(Compute)\\+512MB(Prefetch)} & \makecell{160MB(Compute)\\+1GB(Prefetch)} \\
HBM capacity & 32 GB & 220 GB \\
HBM bandwidth & 1.64 TB/s & 7.4 TB/s \\
\bottomrule
\end{tabular}
}
\label{table_hardware}
\renewcommand{\arraystretch}{1.0}
\end{table}

\begin{table}[t]
\caption{Evaluation datasets\cite{sarathi2}.}
\label{tab:datasets}
\centering
\renewcommand{\arraystretch}{1.5}  
\resizebox{\linewidth}{!}{%
\begin{tabular}{lcccccc}
\toprule
\textbf{Dataset} & \multicolumn{3}{c}{\textbf{Prompt Tokens}} & \multicolumn{3}{c}{\textbf{Output Tokens}} \\
 & Median & P90 & Std. & Median & P90 & Std. \\
\midrule
\makecell[l]{openchat\_\\sharegpt4} & 1730 & 5696 & 2088 & 415 & 834 & 101 \\
\makecell[l]{arxiv\_\\summarization} & 7059 & 12985 & 3638 & 208 & 371 & 265 \\
\bottomrule
\end{tabular}
}
\label{table_dataset}
\renewcommand{\arraystretch}{1.0}
\end{table}

\begin{figure*}[t]
\centerline{\includegraphics[width=38pc]{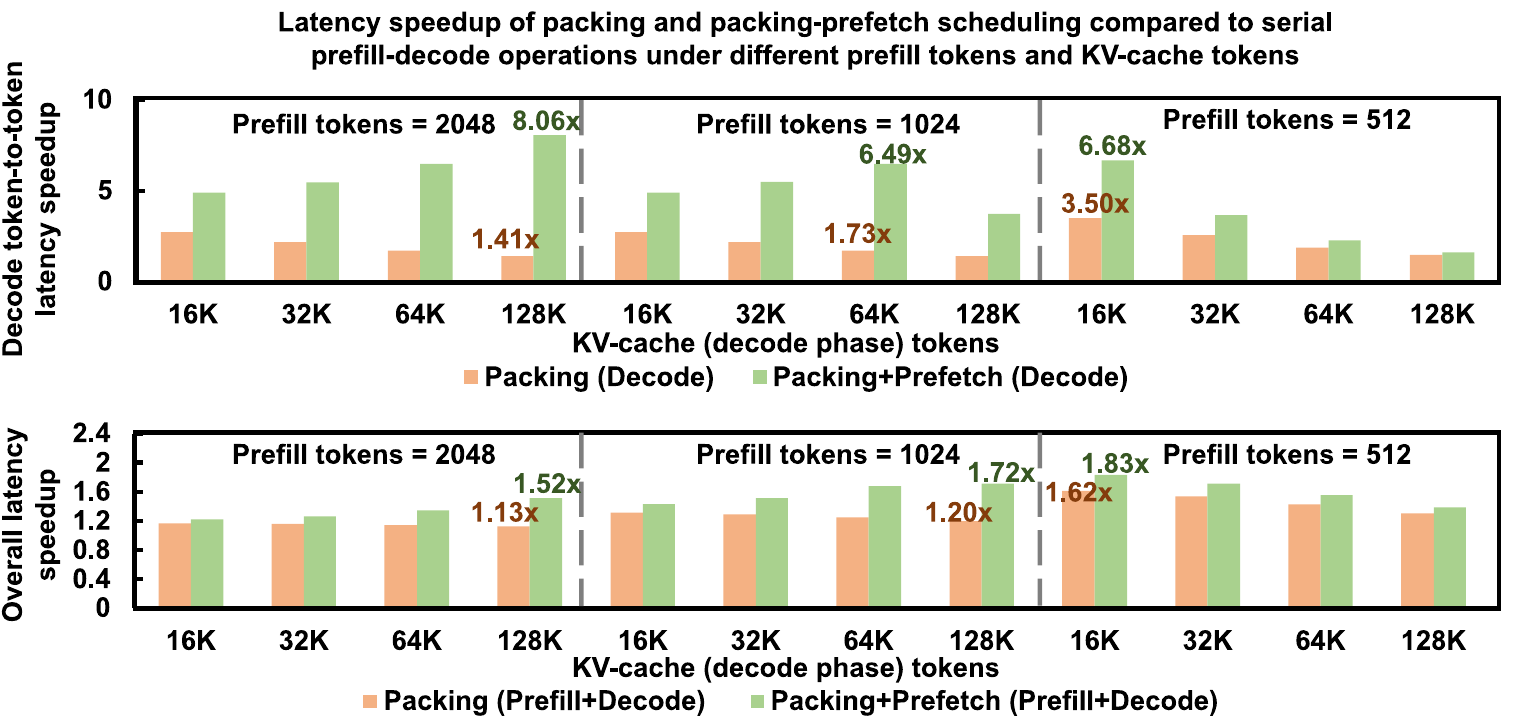}}
\caption{Latency speedup of packing and packing-prefetch scheduling under different prefill tokens and KV-cache tokens on Llama3.1-8B.}
\label{casestudy1}\vspace*{-5pt}
\end{figure*}

\section{Evaluations}

We evaluate the proposed packing-prefetch scheduling optimization on two LLM models and hardware configurations: Llama3.1-8B on a TPUv6e-like architecture and Llama3.1-70B on a TPUv7-like architecture, as shown in Table \ref{tab:llm_hw_config}. Each experiment is conducted on a single processing node under FP16 inference. The on-chip memory for each setup includes two components: (1) an 80MB buffer for the storage of input, weight, and intermediate operand, and (2) a prefetch buffer up to 512MB for Llama3.1-8B and 1GB for Llama3.1-70B. These additional buffer sizes are sufficient to prefetch a single layer’s KV-cache data for one batch of the maximum 128K context-length tokens in the models. Similarly, the HBM capacities on both systems are configured to hold the KV-cache for one batch of 128K tokens in the overall models. We conduct four case studies: case study 1 and 2 focus on stage-level latency analysis, evaluating both Time-between-Tokens latency (TBT) in the decode phase and end-to-end prefill-decode latency, which correlates directly with overall throughput. Case study 3 and 4 evaluate service-level performance using the same real-world multi-request datasets as the prior work\cite{sarathi2}: openchat\_sharegpt4 represents interactive user conversations with ChatGPT-4\cite{gpt4}, arxiv\_summarization features long-form summarization tasks using arXiv.org articles. We demonstrate the throughput improvement in terms of queries per second (QPS) within a 99th percentile (P99) TBT latency constraint and HBM bandwidth savings of proposed approach with both datasets tested on two models. We also explore the throughput-latency tradeoff under two packing chunk sizes on the arxiv\_summarization dataset. 

\textbf{Case study 1}: We first evaluate stage-level latency improvements for the decode operations and the combined prefill-decode packed stage under different combinations of prefill token lengths and KV-cache token sizes using Llama3.1-8B model. Figure \ref{casestudy1} presents the results, where the latency is normalized to the baseline of sequential execution (i.e., prefill followed by decode without packing or prefetching) and reported as speedups. The packing-only baseline uses 80MB of on-chip memory, while our optimization leverages an additional 512MB of M3D memories to store prefetched KV-cache data. For the decode phase, our packing-prefetch scheduling achieves a maximum speedup of 8.06× with 2048 prefill tokens and 128K KV-cache tokens, outperforming the packing-only baseline with a 1.41× speedup under the same conditions. The benefit of prefetching is tied to the available residual HBM bandwidth during prefill. When prefill tokens are lowered to 512 tokens, the residual bandwidth decreases. As a result, speedup drops as KV-cache size increases since the prefetch cannot fully hide the memory latency of decode attention. When considering the overall packed stage latency, the speedups are lower than in the decode case since prefill latency is not optimized and is included in the total runtime. With 512 prefill tokens and 16K KV-cache tokens, our method achieves a 1.83× overall speedup. Additionally, under 1024 prefill tokens, the overall speedup reaches 1.72× compared to 1.20× for the packing-only baseline. While longer prefill sequences provide more bandwidth for KV-cache prefetching and boost decode performance, prefill latency is also increased. Balancing both factors, 1024 prefill tokens show as an effective point in this case study, offering considerable improvements in both decode efficiency and overall runtime.

\begin{figure*}[t]
\centerline{\includegraphics[width=38pc]{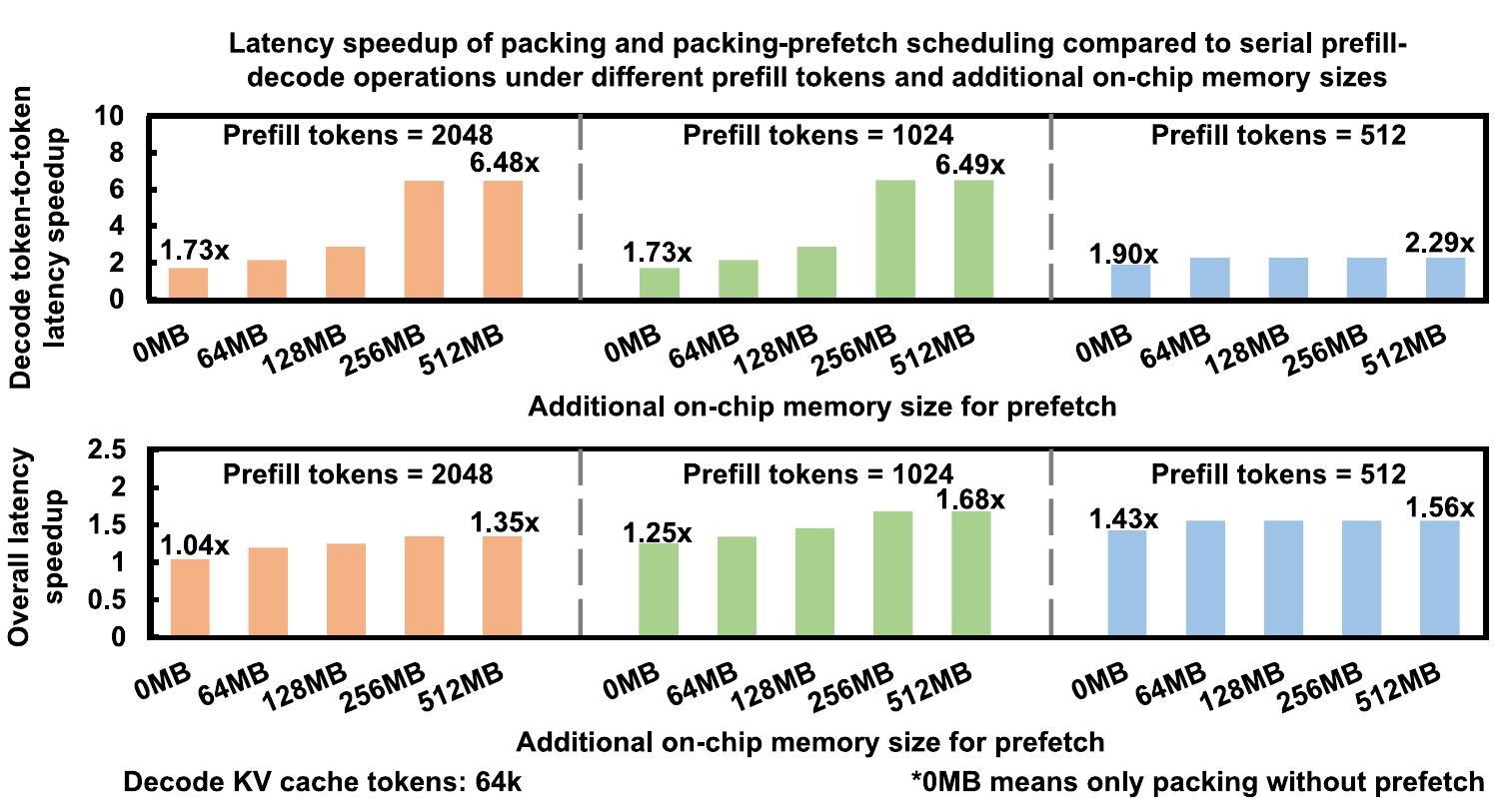}}
\caption{Latency speedup of packing and packing-prefetch scheduling compared to serial prefill-decode operations under different prefill tokens and additional on-chip cache sizes on Llama3.1-8B.}
\label{casestudy2}\vspace*{-5pt}
\end{figure*}

\textbf{Case study 2}: We also explore how varying on-chip memory capacity influences the decode TBT and overall packed latency. Figure \ref{casestudy2} shows the latency speedups of packing-only and packing-prefetch scheduling under different prefill tokens and on-chip memory sizes, with 64K decode KV-cache tokens. Here, 0MB means that no additional on-chip memory is available for prefetching, and only the base 80MB buffer is used for computation, referring to the packing-only baseline. For 2048 and 1024 prefill tokens, we observe similar decode speedups at the same memory size, indicating that prefetching time is sufficient for both cases and speedup is limited by available on-chip memory. As the on-chip memory capacity increases from 0MB to 512MB, the decode TBT speedup improves from 1.73× to 6.49×. For 512 prefill tokens, the speedup improvement is smaller with enlarged memory capacities. This aligns with the findings in case study 1: the shorter prefill phase leaves insufficient time for prefetching regardless of available memory, resulting in limited performance gains. The overall latency speedups follow a similar trend in the decode case. When we increase prefill tokens from 512 to 2048, the speedup of packing-only baseline drops to near 1.0x. With proposed packing-prefetch optimization and the integration of M3D memories, the overall speedup can be improved to 1.35x for 2048 prefill tokens and 1.68x for 1024 prefill tokens.

\begin{figure}
\centerline{\includegraphics[width=21pc]{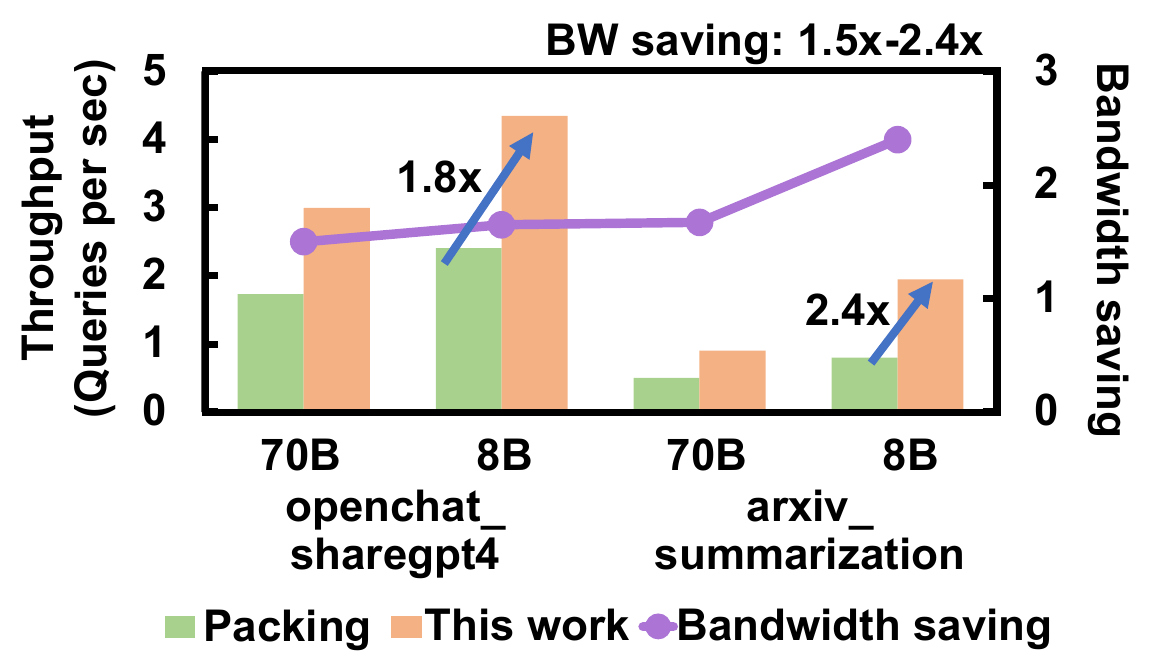}}
\caption{Service-level throughput comparison for two datasets executed on Llama3.1-8B and Llama3.1-70B models and bandwidth savings of packing-only baseline using packing-prefetch scheduling (this work).}
\label{casestudy3}\vspace*{-5pt}
\end{figure}

\textbf{Case study 3}: In addition to stage-level analysis, we evaluate the service-level performance of our proposed packing-prefetch scheduling using two real-world datasets: openchat\_sharegpt4 and arxiv\_summarization, tested on both Llama3.1-8B and Llama3.1-70B models. We adopt the same multi-request packing strategy as used in prior work\cite{sarathi2}, which prioritizes current decode requests and interleaves with the prefill chunk of the requests in the execution queue. The same request arrival time follows Poisson distribution\cite{sarathi2}. To avoid multiple requests pile up in the queue, we set the 99th percentile (P99) request scheduling delay to 1 second. Throughput is measured under a Service-Level Objective (SLO) constraint defined by P99 TBT latency. The latency constraint is evaluated in the condition of 32 concurrent decode requests, each including 4K KV-cache tokens. Based on the evaluation using proposed framework, the SLO thresholds are set to 16.70 ms for Llama3.1-8B and 19.23 ms for Llama3.1-70B. To meet these constraints, the packing chunk size is fixed at 512 tokens. Note that after the prefill phase, the initial number of KV-cache tokens used for decoding matches the number of input prompt tokens. Since each execution cycle can pack multiple decode-phase requests, the total KV-cache size can range from 10K to 128K tokens.

The results are shown in Figure \ref{casestudy3}. First, Llama3.1-8B consistently shows higher throughput than Llama3.1-70B for both datasets. This is because the larger 70B model demands more KV-cache prefetching time with similar improvement of compute performance and HBM bandwidth from TPUv6e-like to TPUv7-like architectures. Second, the throughput of openchat\_sharegpt4 dataset is higher than that of arxiv\_summarization, as the summarization workload involves longer input and output sequences, requiring more compute and data transfer time. Third, despite lower absolute throughput, arxiv\_summarization demonstrates more relative throughput gains from the proposed optimization. For the 8B model, throughput on arxiv\_summarization and openchat\_sharegpt4 improves by 2.4× and 1.8×, respectively. This is attributed to longer prefill tokens in the summarization task, meaning more prefill chunks to be packed with decode requests. In this case, packing efficiency is improved with increased KV-cache prefetching opportunities.

We also compare the packing-only baseline with scaled-up HBM bandwidth to determine how much additional memory bandwidth would be required to match the throughput of our method. The results show that our packing-prefetch scheduling saves up to 2.4x HBM bandwidth under 1.95 QPS for the packing-only baseline in the summarization dataset on Llama3.1-8B. This case study shows that our proposed scheduling optimization improves real-world, long-context service-level inference throughput and alleviates HBM bandwidth pressure.

\begin{figure}[t]
\centerline{\includegraphics[width=21pc]{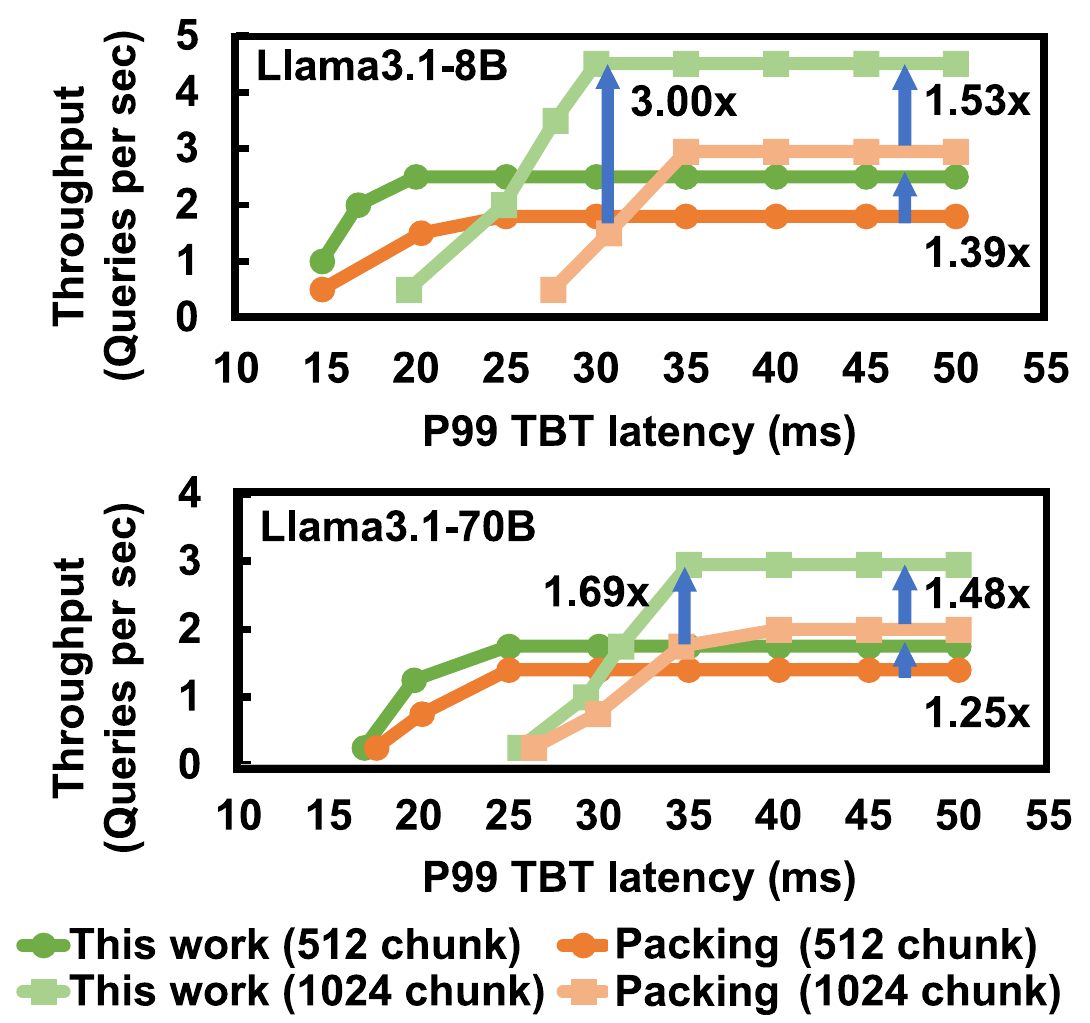}}
\caption{Throughput and latency tradeoff of packing and packing-prefetch scheduling (this work) with different chunk sizes on arxiv\_summerization dataset.}
\label{casestudy4}\vspace*{-15pt}
\end{figure}

\textbf{Case study 4}: We further explore the throughput and latency tradeoff for both packing and packing-prefetch strategies, using chunk sizes of 512 and 1024 tokens on the arxiv\_summarization dataset. This analysis demonstrates how chunk granularity impacts scheduling efficiency under real-world conditions. Similar to the observation in case study 3, Llama3.1-8B achieves higher overall throughput than Llama3.1-70B due to the difference of prefetching time requirement. The throughput saturates with relaxed SLO contraints because the waiting queue fills with more requests' arrival and the system violates scheduling delay constraints. After saturation, our proposed optimization achieves a 1.53× and a 1.39x throughput improvement 1024 chunk size and 512 chunk size, respectively, on Llama3.1-8B. The larger chunk size enables longer prefill windows, leaving more time for KV-cache prefetching of packed decode requests in one execution. This leads to higher peak throughput of 1024 chunk size over 512 chunk size. When applying a tight P99 TBT constraint of 31 ms for the Llama3.1-8B model, our packing-prefetch scheduling shows up to a 3.0× throughput improvement compared to the packing-only baseline.

\section{Conclusion}
In this work, we propose a packing-prefetch scheduling architecture integrated with M3D BEOL memories to address the memory-bound attention bottleneck in the decode phase for long-context LLM inference. Our approach maximizes prefetching benefits by interleaving prefill and decode phases to provide sufficient prefetching time, while leveraging ultra-large capacity on-chip memory to ensure sufficient prefetching space. We evaluate our optimization using a self-developed, end-to-end LLM inference framework. Experiments on Llama3.1-8B and Llama3.1-70B, executed on TPUv6e- and TPUv7-like architectures, demonstrate up to 8.06× reduction in decode latency and 1.83× improvement in overall latency compared to the serial execution. Service-level benchmarks of real-world datasets show 1.7×–2.4× throughput gains and 1.5×–2.4× HBM bandwidth savings over the packing baseline. This work highlights the co-design of scheduling, memory subsystems, and hardware architecture for efficient long-text LLM inference, providing insights to alleviate HBM bottlenecks and bridge the performance gap between on-chip compute and off-chip memory for AI accelerators.
\vspace*{-2pt}

\section*{Acknowledgment}
This work was supported by the Logic Pathfinding Lab of the Samsung Semiconductor.

\vspace*{-8pt}

\end{document}